\documentclass[
  ,draft            
  ,numberedheadings 
  ]
  {aipproc}

\layoutstyle{6x9}
\begin{document}
\title{}
\title[de Sitter equilibrium as a fundamental framework for cosmology]{de Sitter equilibrium as a fundamental framework for cosmology\footnote{To
  appear in the proceedings of the {\em DICE2008} conference, Thomas
  Elze ed..}}  


\author{Andreas Albrecht}{
  address={University of California at Davis\\ Department of
  Physics\\ One Shields Avenue \\ Davis, CA 95616}
}

\begin{abstract} 
Cosmology might turn out to be the study of 
fluctuations around a ``de Sitter equilibrium'' state. In this article
I review the basic ideas and the
attractive features of this framework, and respond to a number common
questions raised about the de Sitter equilibrium picture.  I show that
this framework does not suffer from the ``Boltzmann Brain'' problem,
and relate this cosmological picture to recent work on the ``clock ambiguity''.
\end{abstract}

\maketitle

\section{Introduction}
Most approaches to theoretical cosmology ultimately require a
framework which can quantify the relative probabilities of different
cosmological scenarios.  For example, the widely held belief that
cosmic inflation gives an account of the initial state of the big
bang which is better or ``more natural'' than models without inflation
can only be substantiated within such a framework.  One could hope
that such a framework would allow the various features of inflation
that seem intuitively appealing to directly enhance the probability
assigned to cosmologies with inflation in a quantifiable way.   

A great many models of cosmic inflation yield a picture called
``eternal inflation'' \cite{Linde:1986fd}  where inflation continues forever in 
(exponentially) increasingly  large regions, periodically reheating in
isolated regions known as ``pocket universes'' which can look
something like the big bang cosmology we observe (for some reviews and
a sample of recent work see
\cite{Linde:2007nm,Guth:2007ng,Garriga:2007wz,Linde:2007fr,Aguirre:2006ak,Bousso:2007nd,Linde:2008xf,DeSimone:2008if,DeSimone:2008bq,Bousso:2008hz,Garriga:2008ks}).     
It has long been 
hoped that the eternal nature of eternal inflation would allow the
predictions to be independent of any initial conditions at the start
of inflation, and thus yield stable results that could be regarded as
the robust predictions of the inflationary cosmology. 

So far things have not turned out this way.  Eternal inflation
typically leads to a variety of types of pocket universes, each
produced in infinite 
quantities. To predict which type of pocket universes are most likely one must
regulate these infinities in some way.  To date, there is no
physically determined (or even universally agreed upon) way to
regulate the infinities, and discussions of this matter usually invert
the question to: ``Which of the various possible regularization schemes
yield predictions that actually match what we observe?'' (Many do
not.)   Posing the question this way may be realistic and even
interesting in the absence of something better. But 
as long as one is framing the problem in these terms
one is, at least for the time being, abandoning the prospect of actually making
concrete predictions from inflation.  If one embraces the string
theory landscape picture the problem is only made greater by an
unspecified multitude of possible inflatons experiencing eternal
inflation.  One might hope that a better understanding of
the underlying physics would eventually regulate the infinities for us
and lead to robust predictions (perhaps holographic
considerations\cite{Albrecht:2002xs,ArkaniHamed:2007ky,Dubovsky:2008rf,Garriga:2008ks}
and/or other approaches to carefully developing the formalism\cite{Bousso:2006ev,Linde:2006nw,Hartle:2007gi,Hartle:2008ng,Bousso:2009mw}
will ultimately help with
this).
Another possibility is that the 
current predicament is a reflection of serious problems with the whole
framework of eternal inflation which will prevent it from ever gaining
predictive power.

This paper is about an alternative framework for assigning
probabilities in cosmology which I currently find more promsing than
eternal inflation.  This framework, which I call the ``de Sitter
Equilibrium'' (dSE) cosmology has been examined in two previous
papers\cite{Dyson:2002pf,Albrecht:2004ke}.  In this article I briefly review the dSE picture
and call out the key reasons I find this framework 
attractive.  I then comment on some of the questions that have come
up regarding dSE since the earlier work appeared.   I also offer a quantitative
estimate that illustrates how the dSE picture can evade the so-called
``Boltzmann Brain'' problem.  I then 
discuss how the dSE framework might relate to my recent work on the
``clock ambiguity'' which at least superficially might
appear to be taking things in an orthogonal direction. 

\section{The dSE framework}
\medskip
\subsection{General Features}

Intuitively, one can think of equilibrium for a system as the state a
system achieves if left without external intervention for an
arbitrarily long time. 
By this standard, de Sitter space can be thought of as the equilibrium
state achieved by any system which obeys Einstein gravity with a
positive cosmological constant $\Lambda$.  This point requires the
cosmological constant to be truly constant, not the 
energy of a false vacuum that would ultimately decay through tunneling
processes\cite{Coleman:1980aw} or other instabilities\cite{Mottola:1984ar,Tsamis:1994ca}. 

If our universe does have a true (perfectly stable) positive cosmological constant the
correct long-term description of the universe would be a state 
closely approximating de Sitter space.  The Hawking temperature of de
Sitter space would create fluctuations which could (very
rarely) be sufficiently large to cause a temporary deviation into a 
state that does not look at all like a classical de Sitter space. 

Under these conditions the field of theoretical cosmology become the
study of the full range of fluctuations out of a de Sitter
background, and the interpretation of these different fluctuations as
possible cosmological scenarios.  One very attractive feature of this
picture is that probabilities are assigned to fluctuations in an
equilibrium state without any reference to ``initial conditions'', a
concept which is meaningless in a system eternally in equilibrium.
For fluctuations in an equilibrium state, it is just the laws of
 physics (the Hamiltonian) that determine the probabilities of various
fluctuations.  Another attractive feature of the dSE picture is that the presence of
a horizon surrounding any observer in de Sitter space leads to
quantitative treatments that appears much more naturally finite, in
contrast to the infinities which seem to fundamentally plague eternal inflation. 

Of course, the main prediction of the dSE picture might be taken to be 
that the universe should be observed in a pure de Sitter equilibrium
state.  The viability of dSE as a framework for theoretical cosmology
depends crucially on one's willingness to impose the thermodynamic arrow of
time as a condition, rather than something that
must predicted as a universal and eternal feature of one's theory.
Many authors (although not all \cite{FeynmanMess,Carroll:2004pn}) have indeed found it
reasonable to use the arrow of time as a
condition\cite{Page:1983uh,Albrecht:2002uz,Banks:2003ta,Bousso:2007kq}. This 
could come about through some sort of 
anthropic argument related to how critical the arrow of time is for
the functioning of observers like us, or it could be a much more
narrowly formulated choice to use the fact that we observe an arrow of
time as a condition to pose conditional probability questions about
the universe.

\subsection{de Sitter Entropy}
An observer in a de Sitter space with positive cosmological constant $\Lambda$
is surrounded by an event horizon of radius $R_\Lambda$ given by
\begin{equation}
  R_\Lambda^{-2} = {\rho_\Lambda \over 3 m_P^2} = {\Lambda \over 3}.
\end{equation}
Throughout we use $\hbar = c= k_B = 1$ and $m_P^2 \equiv l_P^2  \equiv
1/8\pi G$.  As with the black hole case, the horizon is associated
with an entropy  
\begin{equation}
S_\Lambda = \pi{R_\Lambda^2 \over l_P^2}
\label{SL}
\end{equation}
Gibbons and Hawking (who were the first to propose and study de Sitter
entropy\cite{Gibbons:1977mu}) showed that when other objects (in particular black
holes) are put in a de Sitter space the overall entropy goes
down. Specifically, when a black hole with entropy $S_{BH}$
is place inside a de Sitter space with entropy $S_\Lambda$ the horizon
of the de Sitter space shrinks so that the entropy of the combined
system is
\begin{equation}
S_\Lambda \rightarrow S_\Lambda - \sqrt{S_\Lambda S_{BH}} + S_{BH}
\approx S_\Lambda - \sqrt{S_\Lambda S_{BH}} .
\label{SLBH}
\end{equation}
Even though the black hole entropy has added to the total, the
decrease due to horizon shrinkage (2nd term) is greater than the
increase due to the additional entropy of the black hole (for black
holes small enough to fit in the de Sitter horizon in the first
place), leading to a net entropy decrease.  Any localized
matter will look like a black hole sufficiently far away and will
decrease the horizon entropy accordingly while adding less entropy
than would be the case for the true black hole.  Similar arguments can
be made about adding a more uniform radiation
field to de Sitter space, and thus the statement that the de Sitter
entropy is the maximum possible entropy for a system with a
stable cosmological constant appears to be quite robust.  This maximal entropy feature is one of the reasons de
Sitter space appears fit to be regarded as an equilibrium state. 

\subsection{Recurrences}
Both previous papers on dSE cosmology have made use of the following
picture:  The de Sitter space is regarded as a finite system in a
Hilbert space of dimension
\begin{equation}
N_\Lambda \equiv e^{S_\Lambda}~.
\end{equation}
One can then make the ergodic argument that a particular fluctuation
which is consistent with $N_F$ microstates occurs with probability
\begin{equation}
P_F = {N_F \over N_\Lambda} \equiv {t_F \over t_R}~.
\label{PF}
\end{equation}
The finite system is expected to experience recurrences on a time $t_R
\propto N_\Lambda$ and the system spends a time $t_F \propto N_F$ in
the fluctuation, which is the origin of the last equality in Eqn. \ref{PF}.
Furthermore, one can infer $N_F$ from the minimum entropy $S_F$ exhibited by 
the entire system during the fluctuation, and thus write
\begin{equation}
P_F = {N_F \over N_\Lambda}\equiv {e^{S_F} \over e^{S_\Lambda}}~.
\label{PF2}
\end{equation}
If the fluctuation is a small localized mass concentration in the
de Sitter background, then (using Eqn. \ref{SLBH})
\begin{equation}
P_F = \exp{\left\{ - \sqrt{ S_\Lambda S_{BH} } \right\}}
\label{PF3}
\end{equation}
where $S_{BH}$ is the entropy of the black hole with equivalent ADM
mass to the localized fluctuation.

\subsection{Probabilities for fluctuation into our observed Universe}
The formalism being outlined here should in principle serve as a means
to calculate the probability of any fluctuation out of the de Sitter
equilibrium.  Of particular interest are the probabilities for a
fluctuation into an inflating state, a fluctuation into a standard big
bang cosmology without inflation, and a fluctuation into a ``Boltzmann
Brain''\cite{Albrecht:2004ke}.  If the probability for a fluctuation to enter
cosmic inflation is much higher than the others, the formalism supports
the prevailing beliefs in theoretical cosmology.  Otherwise, the
formalism is in conflict with these beliefs.  In what follows, we will
simply assume that the physical degrees of freedom include an inflaton
and other fields suitably chosen and coupled to allow the standard
cosmological picture: 
{\em inflation} $\rightarrow$ {\em reheating} $\rightarrow$ {\em
  standard big bang} (SBB) to be a possible behavior of the system.

\subsubsection{Probabilities for the Farhi Guth Guven process}
\label{ASsect}
In \cite{Albrecht:2004ke} Sorbo and I (AS) studied the formation of an inflating
universe form dSE via the following process: A small seed fluctuation of
localized matter forms with probability $P_c$ which then has
probability $P_q$ of fluctuating further into an inflating state.  We
used Eqns. \ref{SLBH}  and \ref{PF3} to calculate
\begin{equation}
P_c = \exp{\left\{-\sqrt{S_\Lambda S_S}\right\}}
\label{PC}
\end{equation}
where $S_S$ is the entropy of the black hole with the same 
ADM mass as the fluctuation. The
(necessarily quantum) probability that the local fluctuation excites
the inflaton field and tunnels into an inflating state is given by 
\begin{equation}
P_q = \exp{\left\{-\pi \left({R_I \over l_P}\right)^2\right\}} \approx exp{\left\{-S_I\right\}}
\label{Pq}
\end{equation}
where $R_I$ is the de Sitter radius during inflation 
\begin{equation}
  R_I^{-2}\equiv {\rho_I \over 3 m_P^2} 
\end{equation}
and $\rho_I$ is the energy density during inflation. This two-stage
process is known as the Farhi Guth Guven process \cite{Farhi:1986ty,Farhi:1989yr}.
Combining Eqns. \ref{PC} and \ref{Pq} gives a probability
\begin{equation}
P_I \equiv P_c P_q = \exp{\left\{-\sqrt{S_\Lambda S_S} - S_I \right\}}
\label{PI}
\end{equation}
for entering into an inflating state.  

One should compare this with the probability of fluctuating into the standard big
bang without inflation.   One can think of the formation of the SBB
without inflation as the time reverse of the process of ``heat death''
of the SBB into de Sitter space.  Seen that way the probability of
fluctuating straight into the SBB is given by equating the entropy of the
fluctuation ($S_F$ in Eqn. \ref{PF2}) to the entropy of the observed universe $S_{SBB} \approx
10^{100}$ (or perhaps one should use $S_{SBB} \approx 10^{85}$, the
entropy the observed universe had in the radiation era).  Either
way, the probability for fluctuating into the SBB without inflation
(from Eqn. \ref{PF2}) is exponentially lower than getting there via
inflation (Eqn. \ref{PI})\footnote{This calculation of
  $P_{SBB}$ follows the calculation of this quantity in
  \cite{Dyson:2002pf}.  In \cite{Albrecht:2004ke} we considered a
  different expression which has a less straightforward motivation
  (and one which I find less compelling at the time I write this)
  One could use either expression for $P_{SBB}$ in what follows
  without changing any of the key points.}
\begin{equation}
{ P_I \over P_{SBB}}  = \exp{\left\{S_\Lambda - \sqrt{ S_\Lambda
    S_{SBB}  }\right\}} \gg 1
\label{PIoPSBB}
\end{equation}
As discussed at length in \cite{Albrecht:2004ke}, this
calculation appears to validate the standard picture of modern
cosmology, where inflation is highly favored over other scenarios.

\subsubsection{DKS probabilities}
\label{DKSsect}
In \cite{Dyson:2002pf} Dyson Kleban and Susskind (DKS) argued that a universe
undergoing inflation looks a lot like a de Sitter space, but with
$\rho_\Lambda = \rho_I$, the energy density of the inflaton field
during inflation. In parallel with Eqn. \ref{SL} one can define
\begin{equation}
S_I = \pi{R_I^2 \over l_P^2} \equiv { \pi \over l_P^2 }  {3 m_p^2 \over \rho_I } 
\label{SI}
\end{equation}
and based on holographic considerations DKS argued that Eqn. \ref{SI}
should give the entropy of the {\em entire universe} during
inflation.  When seen in that light, the probability for a fluctuation
leading to inflation is given by
\begin{equation}
P_I = \exp{ \left\{-\left(S_\Lambda - S_I\right)\right\}} 
\label{PIdks}
\end{equation}
giving
\begin{equation}
{P_I \over P_{SBB}}  = \exp{ \left\{-\left(S_{SBB} -  S_I\right)\right\}} \ll 1
\label{PIoPSBBdks}
\end{equation}
Since inflation is disfavored in this picture it runs totally against
the modern picture of cosmology.  Furthermore DKS argued that, even
ignoring inflation there are lots of cosmologies that would be
exponentially favored over the observed one (for example, cosmologies
with a slightly higher cosmic microwave background temperature and
thus a larger value $S^\prime_{SBB} > S_{SBB}$).

\section{Issues and discussion}

In the this and subsequent sections I will comment on some issues that
have been raised 
in various informal conversations that have taken place since the
publication of \cite{Albrecht:2004ke}. Many are related to the differences
between the AS and DKS calculations outlined in section \ref{ASsect} and
\ref{DKSsect}, so that is a good place to start.

\subsection{dSE as a heat bath}
\label{dSEHB}
In the AS approach the de Sitter space is seen as a large system
for which the fluctuation involves only a small fraction of the entropy.
When the seed forms and fluctuates into an inflating universe, the
full physical system describes {\em both} the de Sitter space
(with the fluctuation present, which just looks like a
small black hole to an observer elsewhere in the de Sitter space) {\em
and} separately the inflating state, destined to 
reheat and form the SBB.   

This picture depends critically on the
belief (not universally shared\cite{Banks:2007ei}) that standard features of
quantum mechanics should extend to quantum gravity so as to allow a
wavefunction that describes one semiclassical spacetime to fluctuate
into one that describes more than one semiclassical spacetime represented
by different physical degrees 
of freedom.  Thus one would expect the wavefunction following the FGG
process to be given by a combination of two 
wavefunctions, one describing the perturbed background de Sitter space
(expressed 
in one subspace) and one describing the inflating state (expressed in
another subspace). As the inflating state reheats, goes through the SBB
phase and eventually achieves ``heat death'' in the final approach to
the background de Sitter space, the two parts of the wavefunctions would
represent the same physical state and the system would again describe only one
semiclassical spacetime, the equilibrium de Sitter space. 

To track the entropy of this process, one must evaluate the entropy of
the entire system (including both semiclassical spacetimes).  From
that perspective, the fact that the actual fluctuation that starts
inflating has a small entropy makes such a fluctuation more likely,
because it removes less entropy from the background de Sitter space
than a larger fluctuation would.  In other words, the entropy of the
entire system makes a smaller drop to start inflation than it would
to fluctuate directly into the SBB cosmology, and that makes inflation
more likely. 

By contrast, the DKS calculation assigns an entropy $S_I$ to the
entire universe once inflation starts.  In that approach, the smaller
$S_I$, the greater the entropic cost of fluctuating into an inflating
state, and generally the cost is much greater to fluctuate into
inflation rather than directly into the SBB. 

At this point whether one chooses the FGG analysis or DKS depends on
what one believes about the way a fundamental theory of quantum gravity
should work (whether, for example, the holographic interpretation of
inflation used by DKS is appropriate).  In this paper I focus on the
FGG process, which results in a cosmological picture which I find quite attractive.

Interestingly, the holographic analysis that leads to Eqn. \ref{PIdks}
in the DKS approach can be duplicated by the
Coleman De Luccia\cite{Coleman:1980aw} (CDL) tunneling process. The CDL process leads to the same
quantitative tunneling probability for entering an inflating state
from the background or ``fundamental'' de Sitter space.  However, if
both the CDL and FGG processes are allowed, FGG will win because it is
much faster\cite{Aguirre:2005nt}.  Some have expressed skepticism that
the FGG process is physically allowed.  If FGG could be eliminated,
then the tunneling analysis will be dominated by CDL and thus give the
same result as the DKS analysis.

\subsubsection{Issues with the FGG process}

Here are a few issues some have with the FGG process, along with my
thoughts in response:

\medskip
{\em Ill-defined path integral:} The path integral methods used by FGG
have some curious properties, including a Euclidean interpolating
solution that is not a manifold.  This has led to some skepticism that
perhaps the FGG process is not valid.  However, Fischler Morgan
and Polchinski (FMP) \cite{Fischler:1989se,Fischler:1990pk} used a Hamiltonian
approach to calculate the same process which does not have any
similar technical peculiarities, and the results were the same (this is
why the FMP approach was used in \cite{Albrecht:2004ke}). 

\medskip
{\em The $m \rightarrow 0$ limit:} In the limit where the mass of the seed
fluctuation that initiates the FGG process is taken to zero, the
process still proceeds at a finite rate. Some cite this as a
problematic feature that suggests the FGG process in unphysical.
Personally, I would expect the $m \rightarrow 0$ behavior
is an artifact of the thin wall approximation (used by both FGG and
and FMP) which is likely to break down at some finite value of the seed
mass (I consider a lower bound on the seed mass based on such
considerations in section \ref{BBsect} below).

\medskip
{\em AdS/CFT calculations:}  Freivogel {\em et al.}
\cite{Freivogel:2005qh} analyzed something like the FGG process using AdS/CFT
techniques. They showed that the FGG process they considered violates unitarity
and thus they argued the FGG process is not physically allowed. However, AdS/CFT analysis is
not able to describe processes that allow the boundary of the bulk to
fluctuate.  While some believe this is a reflection of the fundamental
nature of quantum gravity, another interpretation is that 
this means that an AdS/CFT analysis cannot give a full account of
quantum gravity.  If one takes that view, then the results in
\cite{Freivogel:2005qh} might a reflection of the limitations of the AdS/CFT
analysis.

\section{dSE and Boltzmann Brains}
\label{BBsect}
I now turn to the question of ``Boltzmann Brains'' in dSE cosmology.
The ``Boltzmann Brain problem'' (first posed in \cite{Boltzmann} and
recently receiving renewed attention\cite{Albrecht:2004ke,Page:2006ys,Banks:2007ei,DeSimone:2008if,MersiniHoughton:2008ui})
refers to the situation where one's cosmological framework assigns a
higher probability to isolated ``observers'' fluctuating briefly out of
equilibrium versus observers which are correlated with large
cosmological (non-equilibrium) states of matter the way we are. 
Boltzmann Brain observers may have (false) memories of planet Earth,
the solar system, galaxies, cosmic microwave background maps etc., but
their destiny is to be immediately reabsorbed in the background
equilibrium state.  Since this is not what we experience, cosmological
frameworks which strongly favor Boltzmann Brain observers are
typically considered to be failures.  

In dSE, the probability assigned to a single Boltzmann Brain observer (BB) is
given by using 
\begin{equation}
  S_{BH} = S_{Br}^S  \equiv \left({m_{Br} \over m_P}\right)^2
\end{equation}
in Eqn. \ref{PF3}, where $S_{Br}^S$ is the entropy of a black hole
with the same ADM mass, $m_{Br}$ as the BB, giving
\begin{equation}
P_{Br} = \exp{\left\{-\sqrt{S_\Lambda S_{Br}}\right\}}
\label{PBR}
\end{equation}
To compare this with Eqn. \ref{PI} for the probability for inflation
one must consider the value of $S_S $, the entropy of a black hole
with ADM mass equal to the seed mass $m_S$ for the FGG process.  While 
the $m_S \rightarrow 0$ gives the dominant process, as discussed
above and in \cite{Albrecht:2004ke}, the thin wall approximation
should break down as one takes this limit.  Here we assume the scale
of the inflaton potential gives a lower bound
on $m_S$ given by 
\begin{equation}
m_S = \rho_IR_I^3 = 0.0013kg \left({ \left(10^{16}GeV\right)^4 \over
  \rho_I }\right)^{1/2}~.
\label{MSL}
\end{equation}
Keeping only the dominant terms gives
\begin{equation}
{  P_{Br} \over P_I} ={\exp{\left\{-\sqrt{S_\Lambda S_{Br}}\right\}} \over
  \exp{\left\{-\sqrt{S_\Lambda S_{S}}\right\}}}~.
\label{PBRoPI}
\end{equation}
Inflation is favored when $S_S < S_{Br}$ which gives
\begin{equation}
  \left({m_{Br} \over 0.0013kg}\right) > \left({ \left(10^{16}GeV\right)^4 \over
  \rho_I }\right)^{1/2}
\end{equation}
which is a condition that is very easily met.  Thus the dSE framework
(using the FGG process) does not appear to suffer from the BB problem.

\section{dSE, entropy and time symmetry}
\medskip
\subsection{The low entropy of inflation-produced SBB's}

From a sufficiently fine-grained point of view, an SBB cosmology that
started with reheating at the end of an inflationary era will always
have much lower entropy than other SBB cosmologies that don't have
that special starting point.  This can be illustrated by the fact that
the time-reverse of an SBB cosmology is highly unlikely to suck all the
energy of the universe into the coherent (and extremely low entropy
form) of a homogeneous rolling scalar field. 

Although this feature is
sometimes cited as a problem for inflation
\cite{Penrose:1988mg,Unruh:1996sf,Hollands:2002xi}, in the
context of dSE cosmology things are just as they should be.  It is a
remarkable property of inflation that it can take a low entropy
(inflating) state and reheat it into an state that appears
sufficiently high in entropy (from a suitably coarse-grained point of
view) to look like an SBB cosmology.  But the fact that the
inflationary path to the SBB actually has much lower fine grained
entropy than the ``typical'' SBB allows it to leave more entropy in
the ``heat bath'', making it a much more likely
fluctuation out of dSE.

\subsection{What about the time reversed process?}

As discussed already in section \ref{dSEHB}, the late time evolution of
our observed universe will eventually re-equilibrate with the eternal
de Sitter space.  Also, since there is only some small probability of
tunneling into inflation in the first place, there will be different parts of
the wavefunction representing the part that tunneled into inflation
and the part that did not.  These different parts will ``re-cohere''
as they all approach (or remain in) the dSE state. 

It seems reasonable to expect that this late-time re-equilibration
would look very much the same whether it was an inflation-produced SBB
or a ``regular'' SBB doing the equilibration.  Furthermore, with
nothing defining a universal time direction for the dSE state, one
could equally well expect the ``tunneling out'' and ``equilibrating
back'' events to happen in either order (a meta-observer with a universal
time arrow would equally often see the
time reversed process, but of course a resident of the corresponding SBB phase
would just see time in the direction of increasing entropy). 

The dSE framework implies that the process of an SBB ``decohering
out'' of the dSE (the time reverse of re-equilibration) should be much
more likely if the SBB state in question is destined (due to extremely
subtle microscopic features) to eventually back into an inflating
state, than if it were a ``regular'' SBB with no period of inflation
at the ``beginning''. But it seems extremely hard to imagine how that
could be the case, since it seems natural for the re-coherence process
to look the same in terms of features that matter to the dynamical
properties at that stage (i.e. localized mass concentrations that
could change the de Sitter horizon). 

My response to this point is that indeed it is hard to see how the
probabilities would work out to be so different for the two cases, but
we should be used to that sort of thing being hard.  A box of gas has
a much lower probability of fluctuating in and out of a gold watch
state than it does fluctuating in and out of a pile of dirt state. But
one would expect the final stages of re-cohering to the equilibrium state (at both
ends of the fluctuation) would look very much the same for the cases
of the gold watch and the dirt.  It is
probably very hard to tell at that stage of the evolution what makes
one fluctuation much more probable than the other. So I believe if one
can tolerate that being a difficult problem for the box of gas, one
should be able to tolerate the equivalent problem for the dSE picture
being at least as hard. 

One can also ask whether there is something fundamentally wrong with
the FGG process since it looks so time-asymmetrical (tunneling at one
end and equilibration at the other).  What place is there for such an
asymmetric process in system with so much time symmetry?  Even the
``watch'' and ``dirt'' fluctuations discussed directly above would be
expected to be fully time-symmetric.

\begin{figure}
\includegraphics[width=3.5in]{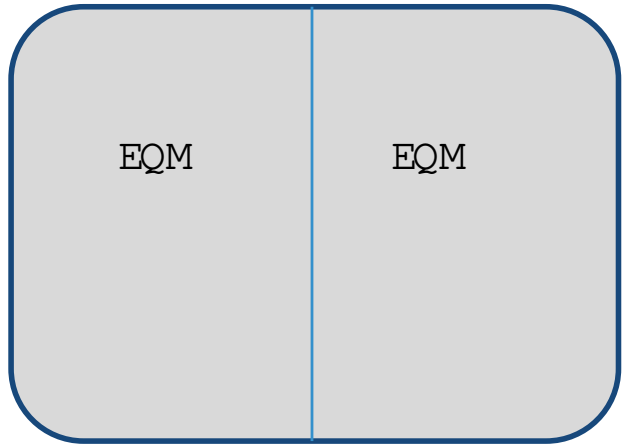}
\caption{\label{Box} A box of gas with a barrier that can only be
  penetrated by quantum tunneling. The gas on both sides is in an
  equilibrium state with the same properties.}
\end{figure}

To further consider this question I propose the following analogy:
Consider a box of gas divided in two by a barrier that can only be
penetrated by quantum tunneling, as depicted in Fig. \ref{Box}. The
gas on both sides of the barrier is in equilibrium, with the same
equilibrium properties on both sides. 
\begin{figure}
\includegraphics[width=3.5in]{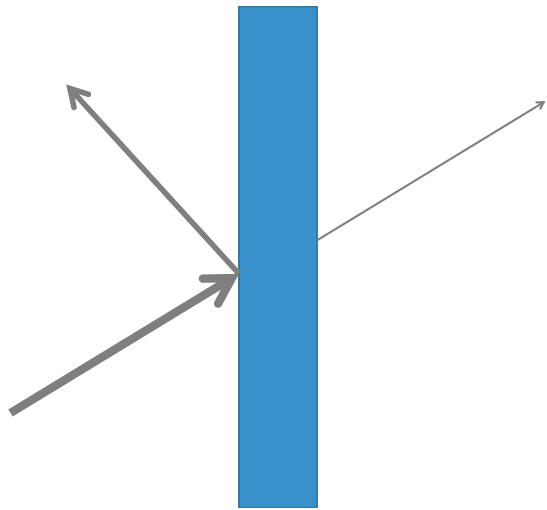}
\caption{\label{Tunnel}A zoomed-in cartoon of a tunneling event at
  the barrier depicted in Fig. \ref{Box}. We find it convenient to
  think of the tunneling in this highly time-asymmetric way, even
  though the system as a whole is time-symmetric.}
\end{figure}

Figure \ref{Tunnel} depicts a zoomed-in view of a gas particle
tunneling through the barrier. The wavefunction of the incoming
particle divides into two pieces, a coherent sum of a piece that
tunnels through and a piece that bounces back.  The time reverse of
this process seems very odd: A particle coherently matched on both
sides of the barrier in order to pull onto just one side after
striking the barrier.  But the box of gas is perfectly
time-symmetrical and the tunneling process occurring in both
directions is equally likely.  I suspect our preference for viewing
this tunneling event in a particular time direction is a
reflection of our own experience living in a very time-asymmetric
world, and I believe a similar assessment applies to concerns about the
apparent time-asymmetry of the FGG process in the dSE picture.

\section{dSE and the Clock Ambiguity}

The dSE cosmological framework appears to be built on laws of physics
that are truly eternal and stable.  One has the picture of a universe
evolving though infinitely many recurrence times and building up
statistics for even the rarest of fluctuations.  Apparently all this
would require perfectly stable physical laws.

In another line of investigation, Iglesias and
I\cite{Albrecht:1994bg,Albrecht:2007mm,Albrecht:2008bj} have studied
a picture motivated by the ``clock ambiguity'' in which the laws of
physics are emergent and are only stable for a finite period of time.
Using arguments developed in \cite{Albrecht:2008bj} (especially Eqn. 8
in that paper) one can find lower bounds on $N_H$, the number of states in the Hilbert
space describing our physical world.  One gets values for
the lower bound around $\exp{\left\{10^{60}\right\}}$ or even
$\exp{\left\{10^{100}\right\}}$, but typically not higher.  It the
context of that work it makes sense to regard the lower bound as the
typical or possibly even the predicted value of $N_H$. From that point
of view, $N_H$ from the clock ambiguity work would be exponentially
smaller than the value $N_H \ge \exp{\left\{S_\Lambda\right\}} \approx
\exp{\left\{10^{120}\right\}}$ required to have stable physical laws
even for one recurrence time.

Thus it would seem that these two ways
of thinking about the cosmos and physical laws (the clock ambiguity
work and the dSE cosmology) are deeply at odds with one another.  That
in itself is not necessarily a problem, since all these ideas lie in
such a speculative domain that this is probably not the right time to
expect them to fit together.  But still, I find it interesting to
reflect a bit further on the possible connections between these ways of
thinking.

In the dSE cosmology, the universe spends by far most of the time
simply sitting in equilibrium, subject to tiny fluctuations that are
completely uninteresting for cosmology.  How much stability is
required of the laws of physics in order to sustain this this
equilibrium picture?  Perhaps not a lot\footnote{This is related to, but not quite the same as a point made in
  \cite{Banks:2002wr} about the stability of quantum measurements over
  a recurrence time
}.  Perhaps these two pictures
can coexist as long as the laws of physics can be stable over
cosmologically interesting timescales, a requirement that we showed in
\cite{Albrecht:2008bj} is quite easy to meet.

\section{Conclusions}
Viewing cosmology as the study of fluctuations
around a de Sitter equilibrium state has a number of appealing
features.  It seems more naturally finite than the picture that
emerges from ``chaotic inflation'' (for which the regulation of
infinities is a problem), and does not depend on what one assumes for
``initial conditions'' (it does not have any).   When viewed from the
point of view of the FGG process (as in \cite{Albrecht:2004ke}), this
framework shows that inflation is favored over big bang cosmologies
without inflation because inflation presents an entropically ``cheaper'' way to
form cosmological fluctuations. In this paper I have also quantified the claim
(first made in \cite{Albrecht:2002uz}) that one of the important roles
of inflation can be to evade the ``Boltzmann Brain'' problem.  I've
also sought to address various questions that have been
raised about this picture, in the hopes of moving those discussions forward. 

There are many open questions:  Which (if either) interpretation of the dSE
cosmology (DKS\cite{Dyson:2002pf} or AS\cite{Albrecht:2004ke}) is
correct? (They give totally different answers.) How can one make the
treatment more rigorous?  (A new formalism for
correlation functions in de Sitter space\cite{Freivogel:2009rf} or
other explorations\cite{Hertog:2005hu} may hold some clues.)  How can the
predictive power be more fully developed?  I find these questions
interesting in their own right, but especially so given the many attractive
features of de Sitter equilibrium cosmology.  Progress on these open
questions will help us see if these ideas can ultimately make good on
their promise to provide deep theoretical foundations for cosmology. 

\begin{theacknowledgments}
  I thank Anthony Aguirre, Raphael Bousso, Alan Guth, Damien Martin and
  Daniel Phillips for discussions that have influenced this work.  Also I thank the organizers,
  especially Thomas Elze, for a really excellent conference.  This
  work was supported in part by DOE grant DE-FG03-91ER40674. 
\end{theacknowledgments}

\bibliographystyle{aipproc}   

\bibliography{DICE}

\IfFileExists{\jobname.bbl}{}
 {\typeout{}
  \typeout{******************************************}
  \typeout{** Please run "bibtex \jobname" to obtain}
  \typeout{** the bibliography and then re-run LaTeX}
  \typeout{** twice to fix the references!}
  \typeout{******************************************}
  \typeout{}
 }

\end{document}